\documentclass{PoS}

\usepackage{amsmath,amssymb}

\usepackage{cite}

\usepackage{comment}

\newcommand{\bff}[1]{\mbox{\boldmath ${#1}$}}
\newcommand{\pvec}{{\mathbf p}}
\newcommand{\qvec}{{\mathbf q}}

\title{Higher-order soft and Coulomb corrections to squark and gluino production at
    the LHC\thanks{Preprint numbers: FR-PHENO-2013-015, SFB/CPP-13-104
TTK-13-25, TUM-HEP-913/13  }}

\ShortTitle{Soft and Coulomb corrections to squark and gluino production at
    the LHC}

\author{Martin Beneke\\

Physik Department T31,
James-Franck-Stra\ss e, Technische Universit\"at M\"unchen,\\
D--85748 Garching, Germany
}
\author{Pietro Falgari\\
Institute for Theoretical Physics and Spinoza Institute,
Utrecht University, \\
3508 TD Utrecht, The Netherlands}

\author{Jan Piclum\thanks{Supported by the DFG Sonderforschungsbereich/Transregio 9 "Computergest\"utzte Theoretische Teilchenphysik"}\\

Physik Department T31,
James-Franck-Stra\ss e, Technische Universit\"at M\"unchen,\\
D--85748 Garching, Germany\\
and\\
Institut f\"ur Theoretische Teilchenphysik und 
Kosmologie,
RWTH Aachen University, \\D--52056 Aachen, Germany
}

\author{\speaker{Christian Schwinn}\\
        Albert-Ludwigs Universit\"at Freiburg, 
        Physikalisches Institut, 
        D-79104 Freiburg, Germany \\
        E-mail: \email{christian.schwinn@physik.uni-freiburg.de}}

      \author{Christopher Wever\thanks{ partially funded by Research
          Funding Program ARISTEIA, HOCTools (co-financed by
          the European Union (European Social Fund ESF) and Greek
          national funds through the Operational Program "Education
          and Lifelong Learning" of the National Strategic Reference Framework (NSRF)) and by the research programme Mozaiek (NWO)}\\
        Institute of Nuclear Physics, NCSR "Demokritos", GR 15310 Athens, Greece   
}

\abstract{We present predictions for the total cross sections for pair
  production of squarks and gluinos at the LHC including a combined
  NNLL resummation of soft and Coulomb gluon effects. The NNLL
  corrections can be up to $25\%$ relative to previous NLL results and
  reduce the theoretical uncertainties to the $10\%$ level.}

\FullConference{11th International Symposium on Radiative Corrections (Applications of Quantum Field Theory to Phenomenology) (RADCOR 2013),\\
		22-27 September 2013\\
		Lumley Castle Hotel, Durham, UK }

\begin{document}

\section{Introduction}
Supersymmetry (SUSY) and its realization in the $R$-parity
conserving Minimally Supersymmetric Standard Model  is a
well-studied and motivated extension of the Standard Model. It could
provide a solution to shortcomings of the SM such as the absence of a
dark matter candidate and  it might stabilize  the electroweak scale
against quantum corrections.
The search for SUSY at the TeV scale is therefore a central part of the physics program of the Large Hadron Collider
(LHC).
At hadron colliders,
the production of squarks and gluinos, the superpartners of quarks and
gluons, is expected to be the dominant signature.  Current LHC  limits
exclude gluino masses up to
$m_{\tilde g}=1.3$~TeV and superpartners of the quarks of the first two generations 
below $m_{\tilde q}\lesssim 800$~GeV. Equal squark and
gluino masses can be excluded up to $m_{\tilde g}\sim 1.7$~TeV~\cite{SUSY-exp}.
However, these bounds depend on assumptions e.g. on the mass of the
lightest supersymmetric particle and can be evaded for instance by
compressed mass spectra. The search for SUSY will therefore remain a focus of the
$13$-$14$~TeV run of the LHC that has the potential to discover or
exclude squarks and gluinos in the $3$~TeV range.
Turning exclusion limits on production cross sections
into bounds on superparticle masses requires precise predictions for
these cross sections. 
In this contribution we report on the
status of predictions for squark and gluino production at
the LHC and present first results of a combined NNLL resummation of soft-gluon and Coulomb corrections~\cite{Beneke:SUSY}.

\section{Squark and gluino production at the LHC}

\begin{figure}[t!]
  \centering
  \includegraphics[width=0.4 \linewidth]{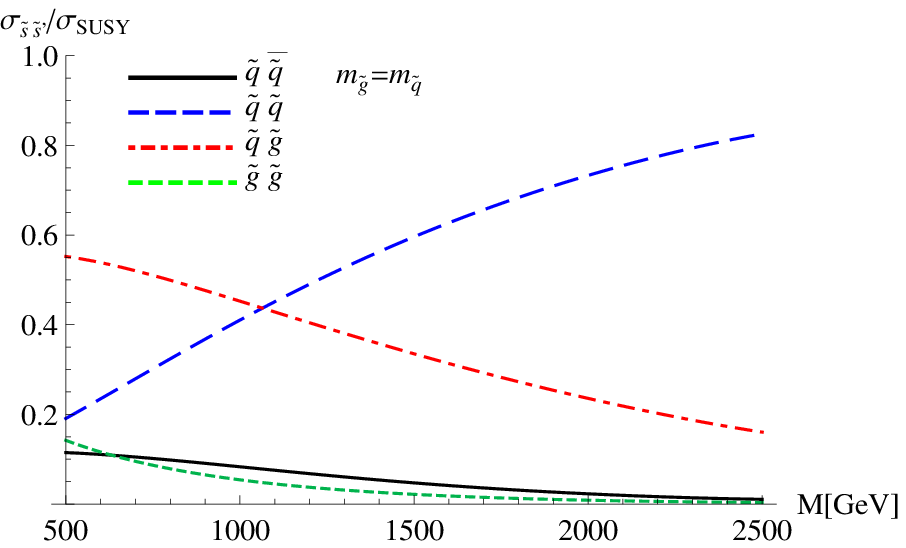}
  \includegraphics[width=0.4 \linewidth]{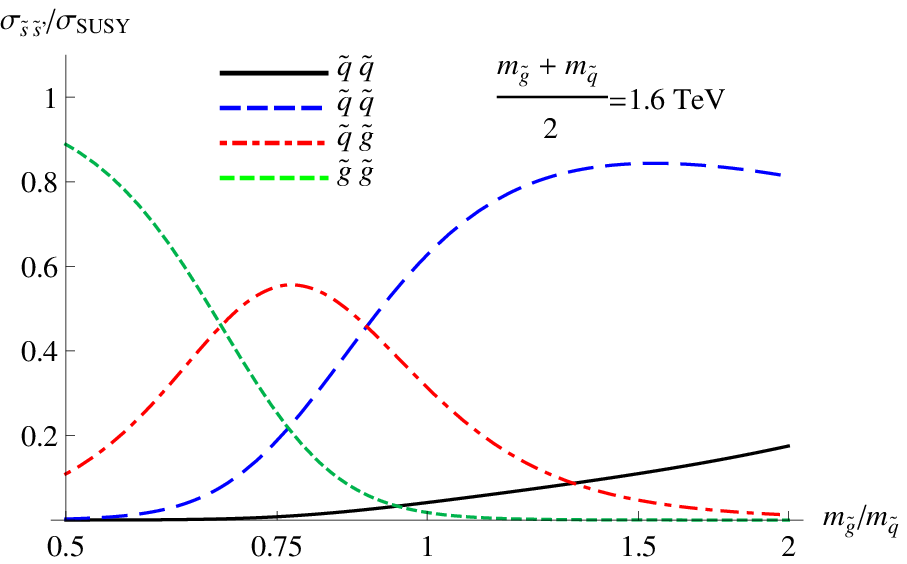}\\
\includegraphics[width=0.4 \linewidth]{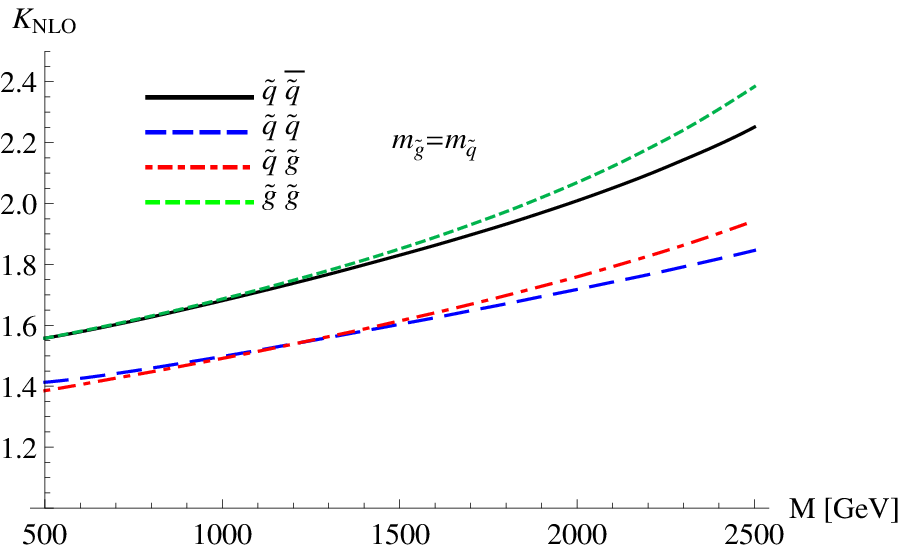}
\includegraphics[width=0.4 \linewidth]{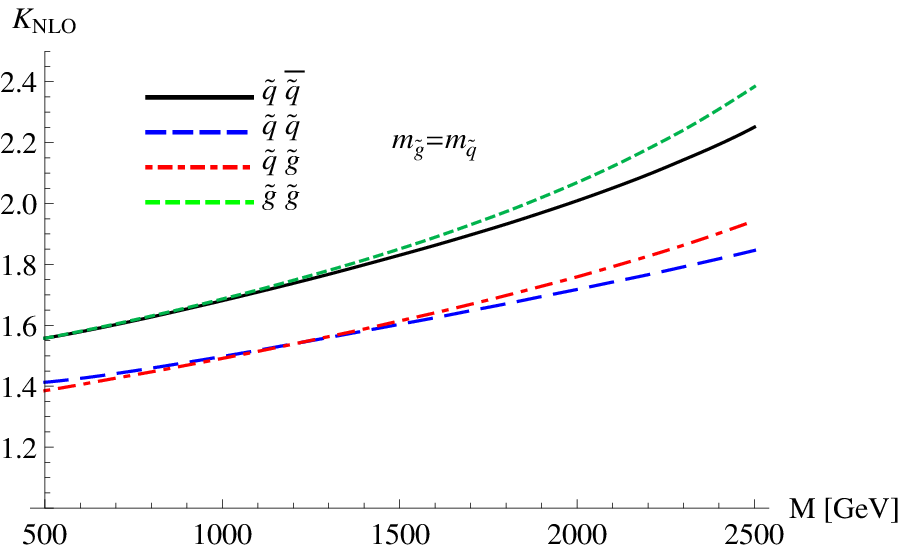}
\caption{Top: Relative contribution of  the
  different squark and gluino production processes 
  to the total Born production rate of coloured sparticles,
  $\sigma_{\text{SUSY}}$, for the LHC with $\sqrt s=8$ TeV. Below: NLO
  $K$-factor for squark and gluino production processes at the LHC.
  The left plots show the mass dependence for $m_{\tilde q}=m_{\tilde
    g}=M$ while the right plot shows the dependence on the ratio
  $m_{\tilde g}/m_{\tilde q}$ for a fixed average mass
  $(m_{\tilde{q}}+m_{\tilde{g}})/2=1.6$~TeV. In the $K$-factors the MSTW2008NLO PDFs~\cite{Martin:2009iq} have been used for the LO and NLO cross sections.}
  \label{fig:LO-NLO}
\end{figure}

At hadron colliders, light-flavour squarks and gluinos, denoted
jointly by $\tilde{s}$, $\tilde{s}' \in\{\tilde q,\bar{\tilde
  q},\tilde g\}$, can be pair-produced through partonic production
processes of the form $ pp'\to \tilde{s} \tilde{s}'X$ from the
incoming partons $p,p'\in\{q,\bar q,g\}$.  The relevant
production channels at leading order~\cite{Kane:1982hw} are given by
\begin{align}
\label{eq:processes}
 gg, \, q \bar{q} &\rightarrow \tilde{q} \bar{\tilde{q}} \,, &
  q q&\rightarrow \tilde{q} \tilde{q},  \, &
  g q &\rightarrow \tilde{g} \tilde{q},  \, &
  gg, \, q \bar{q}  & \rightarrow \tilde{g} \tilde{g}  \,, 
\end{align}
and the corresponding charge-conjugated channels for squarks. 
Flavour indices of quarks and squarks have been suppressed. For  the light-flavour squarks 
 a common mass $m_{\tilde{q}}$ will be used.
In this contribution we will not consider the production of stop
pairs that has been discussed in~\cite{Falgari:2012hx}.  In
the upper plots in Figure~\ref{fig:LO-NLO} the relative contribution of the
processes~\eqref{eq:processes} to the inclusive  squark and
gluino production cross section $\sigma_{\text{SUSY}}=\sigma_{
  \tilde q\bar{\tilde q}+ \tilde q\tilde q+ \tilde g \tilde q+ \tilde
  g\tilde g}$ is shown for the LHC with $\sqrt s=8$~TeV centre-of-mass
energy.  The left-hand  plot displays the relative contributions
of the processes~\eqref{eq:processes} as a function of a common squark and gluino
mass, while in the right-hand  plot the relative contributions are
shown as a function of the squark-gluino mass ratio.
The results for the $K$-factor
$K_{\text{NLO}}=\sigma_{\text{NLO}}/\sigma_{\text{LO}}$ for the  NLO SUSY-QCD corrections~\cite{Beenakker:1996ch} obtained with the program
\texttt{PROSPINO}~\cite{Beenakker:1996ed}  in the
lower plots in Figure~\ref{fig:LO-NLO}  show that  the NLO  corrections can be of the order of $100\%$ of
the tree-level cross section.

The large NLO K-factors 
 can be attributed  to the enhancement of
radiative corrections in the threshold limit $\beta \equiv \sqrt{1-4
  M^2/\hat{s}} \rightarrow 0$, with  the average sparticle mass $M=\frac{1}{2}(m_{\tilde
  s}+m_{\tilde {s'}}) $  and the partonic centre-of-mass energy  $\hat{s}$. In this limit
the partonic cross section is dominated by soft-gluon emission off the
initial- and final-state coloured particles and by Coulomb
interactions of the two non-relativistic heavy particles, which give
rise to singular terms of the form $\alpha_s \ln^{2,1} \beta$ and
$\alpha_s/\beta$, respectively.  The radiative corrections in
the threshold limit can be written in a simple and process-independent form 
using a colour decomposition of the total partonic cross section,
\begin{equation} \label{eq:partonic}
 \hat{\sigma}_{p p'}(\hat s,\mu_f)= \sum_{R_\alpha} \hat{\sigma}^{(0),R_\alpha}_{p p'} (\hat s,\mu_f)
\left\{1+\frac{\alpha_s}{4 \pi} f^{(1),R_\alpha}_{p p'} (\hat s,\mu_f)+... \right\} \, ,
\end{equation}
where $\mu_f$ is the factorization scale,  
$R_\alpha$ are the irreducible representations  in the
decomposition $R\otimes R'=\sum R_\alpha$ of the product of the
$SU(3)$ representations $R$ and $R'$ of the two final-state
sparticles and $\hat{\sigma}^{(0),R_\alpha}_{p p'}$  are the Born
cross sections  projected on  the colour channel
$R_\alpha$~\cite{Kulesza:2008jb,Beenakker:2009ha}. The relevant decompositions for squark and gluino
production are given by
\begin{equation}
\label{eq:susy-reps}
  \begin{aligned}
\tilde q\bar{\tilde q}&:&3\otimes \bar 3&=1\oplus 8\,,&
\tilde q \tilde q&:& 3\otimes 3&= \bar 3 \oplus 6\,,\\ 
 \tilde q\tilde g&:&3\otimes 8&=3\oplus \bar 6 \oplus 15\,,&
  \tilde g\tilde g&:&   8 \otimes 8&=1 \oplus 8_s \oplus 8_a \oplus 10
  \oplus \overline{10} \oplus 27\,.
  \end{aligned}
\end{equation}
The NLO scaling functions $f^{(1),R_\alpha}_{p p'} $ assume a simple form in
the threshold limit~\cite{Beneke:2009ye}:
\begin{eqnarray}
\label{eq:NLOapprox}
f^{(1),R_\alpha}_{p p'}(\hat s, \mu_f) &=& -\frac{2 \pi^2 D_{R_\alpha}}{\beta} \sqrt{\frac{2 m_r}{M}}+4 (C_r+C_{r'}) \left[\ln^2 \left(\frac{8 M \beta^2}{\mu_f}\right)+8-\frac{11 \pi^2}{24}\right]\nonumber\\
&&-4 (C_{R_\alpha}+4 (C_r+C_{r'})) \ln \left(\frac{8 M \beta^2}{\mu_f}\right)+12 C_{R_\alpha}+h_{pp'}^{(1),R_\alpha}+{\cal O}(\beta).
\end{eqnarray}
Here $m_r=m_{\tilde{s}} m_{\tilde{s}'}/(m_{\tilde{s}}+m_{\tilde{s}'})$
is the reduced mass, $r$ and $r'$ are the colour representations of
the initial partons $p$ and $p'$, and $C_{R}$ is the quadratic Casimir
invariant for a representation $R$.  The Coulomb coefficients for
sparticles in the representations $R$ and $R'$ in the colour channel
$R_\alpha$ read $ D_{R_\alpha}=\frac{1}{2}(C_{R_\alpha}-C_R-C_{R'}) $
where negative (positive) values correspond to an attractive
(repulsive) Coulomb potential.  The process-dependent coefficients
$h_{pp'}^{(1),R_\alpha}$ have been recently computed for all squark
and gluino production
processes~\cite{Beenakker:2011sf,Kauth:2011vg,Langenfeld:2012ti,Beenakker:2013mva}.
The singular threshold corrections, i.e. all terms
in~\eqref{eq:NLOapprox} apart from constants, usually dominate the
total NLO cross section and differ from the exact NLO result by
typically only $10-30\%$~\cite{Falgari:2012hx}.  This motivates the
computation of the higher-order threshold-enhanced terms, as discussed
in the remainder of this contribution.

%%%%%%%%%%%%%%%%%%%%%%%%

\section{Joint soft and Coulomb resummation}
\label{sec:resum}

Near the partonic production threshold $\beta\to 0$ the conventional perturbative expansion in $\alpha_s$ breaks down and the perturbative series has to be rearranged by treating both threshold logarithms $\alpha_s
\ln\beta$ and Coulomb corrections $\alpha_s/\beta$  as quantities of order one.
The accuracy of the rearranged perturbative series  can be defined by representing the
resummed cross section schematically as
\begin{equation}
\label{eq:syst}
\hat{\sigma}_{p p'} = \,\hat \sigma^{(0)}_{p p'}\, 
\sum_{k=0}^\infty \!\left(\tfrac{\alpha_s}{\beta}\right)^{\!k} \,
\exp\Big[\underbrace{\ln\beta\,g_0(\alpha_s\ln\beta)}_{\mbox{(LL)}}+ 
\underbrace{g_1(\alpha_s\ln\beta)}_{\mbox{(NLL)}}+
\underbrace{\alpha_s g_2(\alpha_s\ln\beta)}_{\mbox{(NNLL)}}+\ldots\Big]
\Bigl\{1+
 \alpha_s c_{\text{NNLL}}+\ldots\Bigr\}\, . 
\end{equation}
Methods for the separate resummation of threshold
logarithms~\cite{Sterman:1986aj,Kidonakis:1997gm,Becher:2006nr}
and Coulomb cor\-recti\-ons~\cite{Hoang:2000yr} are well
known. Applications to squark and gluino
production include NLL resummation of threshold
logarithms~\cite{Kulesza:2008jb,Beenakker:2009ha,Beneke:2010da},
Coulomb resummation~\cite{Kulesza:2008jb,Hagiwara:2009hq,Beneke:2010da,Kauth:2011vg},
 approximate NNLO
calculations~\cite{Langenfeld:2009eg,Langenfeld:2012ti} and 
NNLL resummation of threshold logarithms  for
some processes~\cite{Beenakker:2011sf,Pfoh:2013iia}.

The combined NLL resummation of Coulomb and soft effects has been
performed for squark-antisquark production in~\cite{Beneke:2010da} and
all other processes in~\cite{Falgari:2012hx}, where it was found that
 Coulomb corrections and soft-Coulomb
interference can be as large as the soft corrections alone.
In the following, we discuss the extension of this result to NNLL.
Up to this accuracy, 
partonic cross sections in the limit $\beta\to 0$  factorize into a
hard function $H^{R_\alpha}$, a soft function $W^{R_\alpha}$ and a
Coulomb function $J_{R_\alpha}$~\cite{Beneke:2009rj,Beneke:2010da}:
\begin{equation}
\label{eq:fact}
  \hat\sigma_{pp'}(\hat s,\mu_f)
= \sum_{R_\alpha}H^{R_\alpha}_{pp'}(m_{\tilde q},m_{\tilde g},\mu_f)
\;\int d \omega\;
J_{R_\alpha}(M\beta^2-\frac{\omega}{2})\,
W^{R_\alpha}(\omega,\mu_f)\, .
\end{equation}
The hard function encodes the
partonic hard-scattering processes and is related to squared on-shell
scattering amplitudes at threshold. The
potential function sums exchange of Coulomb gluons associated
to corrections $\sim (\alpha_s/\beta)^n$ while the soft function sums
the threshold logarithms.  The convolution of the soft- and potential
functions accounts for the energy loss of the squark/gluino system due
to soft gluons with energy of the order $M\beta^2$.
Near threshold, soft-gluon radiation is only sensitive to the total colour
state $R_\alpha$ of the non-relativistic squark/gluino system, as has been
 shown to all orders in the strong coupling~\cite{Beneke:2009rj}, consistent 
with explicit one-loop
calculations~\cite{Kidonakis:1997gm,Kulesza:2008jb,Beenakker:2009ha}.
  The formula~\eqref{eq:fact} has been derived for particles
dominantly produced in an $S$-wave, i.e.\ with a cross section $\hat
\sigma\sim \beta$, which is the case for all  production channels of light-flavour squarks and gluinos, and for processes with a  leading $P$-wave contribution  $\hat \sigma\sim \beta^3$~\cite{Falgari:2012hx}, as for stop-antistop production from a quark-antiquark initial state.

Resummation of threshold logarithms is performed by evolving the soft
function from a soft scale $\mu_s\sim M\beta^2$ to a hard-scattering
scale $\mu_f\sim M$ using a renormalization-group equation derived
in~\cite{Beneke:2009rj} with results
from~\cite{Becher:2009kw}  (equivalent results have been obtained
independently  in
the traditional Mellin-space approach~\cite{Czakon:2009zw}). The hard
function is evolved
from a scale $\mu_h\sim 2M$ to $\mu_f$. 
In the momentum-space
formalism~\cite{Becher:2006nr} 
the resummed cross section can be written as 
\begin{align}
\hat\sigma^{\text{res}}_{pp'}(\hat s,\mu_f)=&
\!\!\sum_{R_\alpha}
H^{R_\alpha}_{pp'}(\mu_h)
 U_{R_\alpha}(\mu_h,\mu_s,\mu_f)\!\!
 \left(\frac{2M}{\mu_s}\right)^{-2\eta} %\\
%&\times
\!\!\!\!\!\!\!
\tilde{s}^{R_\alpha}(\partial_\eta,\mu_s)
\frac{e^{-2 \gamma_E \eta}}{\Gamma(2 \eta)}\!\!
\int_0^\infty \!\!\!\!\! d \omega 
\frac{ J_{R_{\alpha}}(M\beta^2-\tfrac{\omega}{2})}{\omega}
\left(\frac{\omega}{\mu_s}\right)^{2 \eta}  \!\!\!\!\!\!\!\!\!.
\label{eq:resum-sigma}
\end{align}
Resummation at NNLL accuracy requires the expansions of the hard function and the
Laplace-transformed soft function~\cite{Beneke:2009rj}  up to  NLO,
\begin{align}
\label{eq:hard-def}
   H^{R_\alpha}_{pp'}(\mu_h)&= H^{R_\alpha(0)}_{pp'}(\mu_h)
   \left[1+\frac{\alpha_s(\mu_h)}{4\pi}h^{R(1)}_{pp'}(\mu_h)
     +\mathcal{O}(\alpha_s^2)\right] \,,\\
 \tilde  s^{R_\alpha}(\rho,\mu)&=\int_{0_-}^{\infty} d \omega e^{-s \omega}
\,  W^{R_\alpha}(\omega,\mu) =1 +\frac{\alpha_s}{4\pi} 
\left[\left(C_r+C_{r'}\right)\left(   \rho^2+\frac{\pi^2}{6}\right)
- 2C_{R}\left( \rho-2\right)  \right]+\mathcal{O}(\alpha_s^2),
    \end{align}
    with $s = 1/(e^{\gamma_E} \mu e^{\rho/2})$.  The one-loop hard
    coefficients are the same as in~\eqref{eq:NLOapprox}.  The functions $U_{R_\alpha}$ and
    $\eta$ contain logarithms of the ratios of the various scales,
the explicit expressions
 at NNLL can  be found in~\cite{Becher:2006nr}. 

For NNLL accuracy, the NLO potential function is required that can be
written as~\cite{Beneke:2011mq}
\begin{equation}
J_{R}(E)=2 \,\mbox{Im} \left[\,
G^{(0)}_{C,R}(0,0;E) \,\Delta_{\rm nC}(E) + G^{(1)}_{C,R}(0,0;E) + 
\ldots\right] \, , 
\label{JRal}
\end{equation}
where $G^{(0)}_{C,R}$ is the solution to the Schr\"odinger equation with the
leading Coulomb potential, resumming all $(\alpha_s/\beta)^n$
corrections. The function $G^{(1)}_{C,R}$ sums $\alpha_s \times (\alpha_s/\beta)^n$
corrections by solving perturbatively the Schr\"odinger equation with one  
insertion  of the NLO Coulomb potential,
\begin{equation}
\delta\tilde{V}(\bff{p},\bff{q}) = 
\frac{4\pi D_{R}\alpha_s(\mu)}{\bff{q}^2}  \frac{\alpha_s(\mu)}{4\pi} 
\left(a_1-\beta_0\ln\frac{\bff{q}^2}{\mu^2}\right)\,,
\label{delV}
\end{equation}
where $\beta_0$ is
the one-loop beta-function coefficient, and $a_1 =\frac{31}{9}
C_A-\frac{20}{9} n_l T_f$. 
The factor $\Delta_{\rm nC}$ arises from non-Coulomb NNLO potential
terms~\cite{Beneke:1999qg}.
For squark and gluino production, these read~\cite{Beneke:SUSY}
\begin{eqnarray}
  \delta\tilde{V}_{\text{\tiny{NNLO}}}(\bff{p},\bff{q})&& =
\frac{4\pi D_{R}\alpha_s(\mu^2)}{\bff{q}^2} 
\left[ \frac{\pi\alpha_s(\mu^2)|\bff{q}|}{8 m_r}
\left(\frac{D_{R}}{2}\frac{2m_r}{M}+C_A \right) \right. \nonumber \\
&& +
\frac{\pvec^2}{m_{\tilde s}m_{\tilde{s}'}} - \frac{\qvec^2}{8m_{\tilde s}^2m_{\tilde{s}'}^2} ( 2m_{\tilde s}m_{\tilde{s}'} +
    m_{\tilde s}^2\, c_2(m_{\tilde{s}'}) + m_{\tilde{s}'}^2\, c_2(m_{\tilde s}) ) \nonumber \\
&& 
\left.    +  \frac{1}{16m_{\tilde s}m_{\tilde{s}'}} [\sigma^i,\sigma^j] q^j \otimes
  [\sigma^i,\sigma^k] q^k  +\dots   \right]\!\!,
\label{delV2}
\end{eqnarray}
where terms not contributing to squark and gluino production processes
are not shown.
For scalars the spin-dependent terms are set to zero. 
The matching 
coefficient $c_2$ has the
 tree-level value zero (one) for scalars (fermions).
Projecting on the relevant spin states, 
the non-Coulomb  correction in~\eqref{JRal} is obtained as
\begin{equation}
\Delta_{\rm nC}(E) = 1+ 
\alpha_s^2(\mu_C)\,\ln\beta \,\left[-2 D_{R}^2 \,(1+v_{\rm spin}) 
+ D_{R} C_A\right] \theta(E),
\end{equation}
where the spin-dependent coefficient for the squark and gluino
production processes is given by
\begin{equation}
  \begin{aligned}
v_{\rm spin}(\tilde q\bar{\tilde{q}})=
v_{\rm spin}(\tilde q\tilde q) &= -\frac{2m_{r}}{4M} ,&
v_{\rm spin}(\tilde q\tilde g)  &=
 \frac{1}{2}\left(\tfrac{m_{\tilde g}^2}{(m_{\tilde q}+m_{\tilde g})^2}-1\right),\\
v_{\rm spin}((\tilde g\tilde g)_{S=0})&= 0  ,&
v_{\rm spin}((\tilde g\tilde g)_{S=1})&=  -\frac{2}{3}.    
  \end{aligned}
\end{equation}
The gluino pairs are
produced with spin $S=0$ for the symmetric 
colour
representations $1$, $8_s$, $27$ and with $S=1$ for
anti-symmetric colour representations 
$8_a$, $10$ (see
e.g.~\cite{Kauth:2011vg}).
In the colour channels with an attractive Coulomb potential, the Coulomb Green function develops bound-state poles below threshold.
We include these bound-state contributions and convolute them with the soft corrections as described in~\cite{Beneke:2011mq}.
If the finite decay width of squarks and gluinos is taken into account, the bound-state poles are smeared out. This has been investigated at NLL accuracy in~\cite{Falgari:2012sq} with the conclusion that for $\Gamma_{\tilde s}/m_{\tilde s}\lesssim 5\%$ the uncertainties due to finite width effects are smaller than the uncertainties of the NLL calculation.

\section{Squark and gluino production at NNLL}
We have implemented the NNLL resummation discussed in
Section~\ref{sec:resum} following the treatment of
top-quark pair production  in~\cite{Beneke:2011mq,Beneke:2012wb}. A public
program based on \texttt{topixs}~\cite{Beneke:2012wb} is in
preparation.  As in the previous NLL
resummation~\cite{Falgari:2012hx}, the LO hard functions are expressed
in terms of the exact colour separated Born cross sections. No
resummation is performed  for colour channels that are suppressed at threshold. The
convolution of the resummed partonic cross section with the PDFs is
regularized as discussed in~\cite{Beneke:2011mq}.  The NNLL
cross section is matched to the sum of the exact NLO
result~\cite{Beenakker:1996ch} from \texttt{PROSPINO} and the
approximate NNLO cross section~\cite{Beneke:2009ye} where double
counting is avoided by subtracting the NNLO-expansion of the resummed cross
section. 
 In order to see the impact
of Coulomb resummation,
 we also consider an approximation
$\text{NNLL}_{\text{fixed-C}}$ where the product of hard and Coulomb
corrections is replaced by its expansion up to
$\mathcal{O}(\alpha_s^2)$.

\begin{figure}[t!]
  \centering
  \includegraphics[width=0.4 \linewidth]{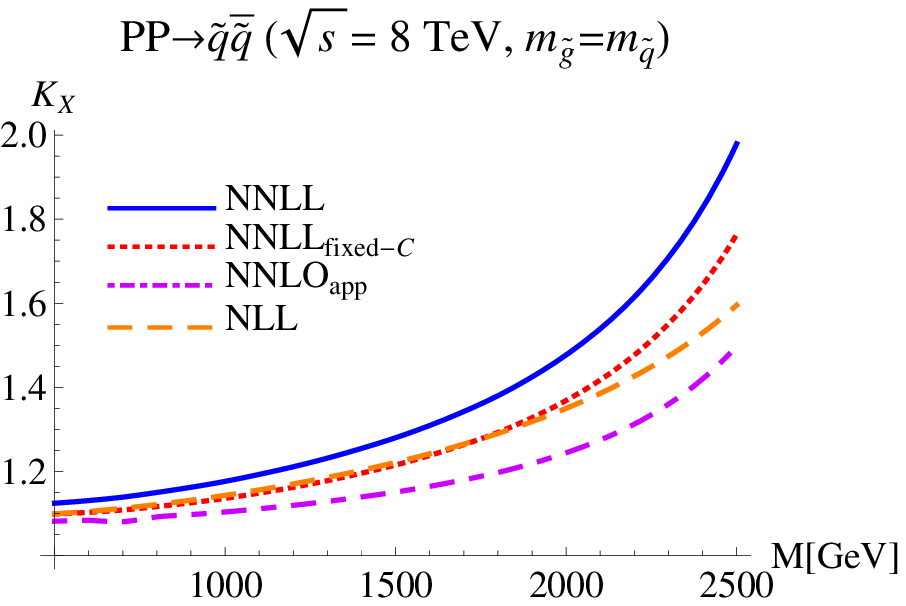}
 \includegraphics[width=0.4 \linewidth]{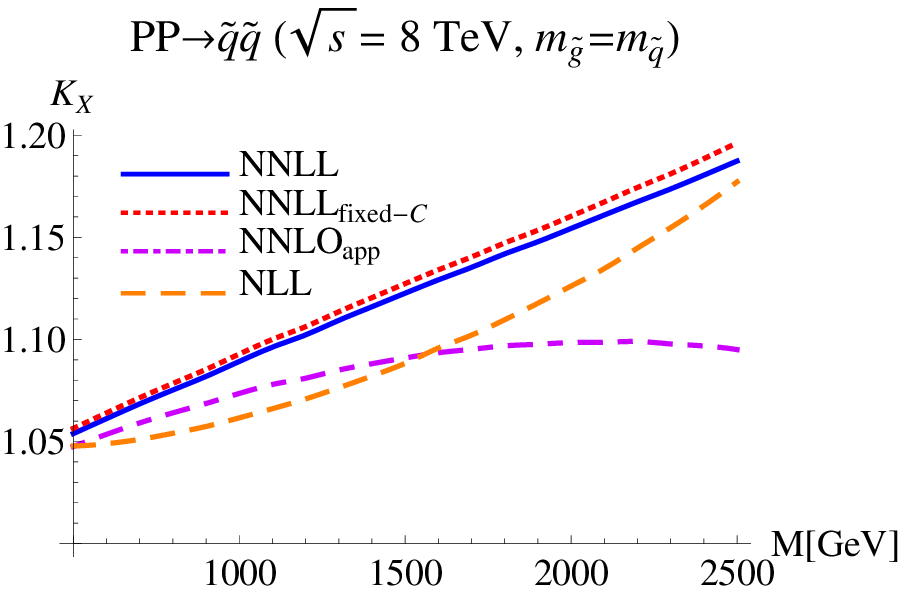}\\
\includegraphics[width=0.4 \linewidth]{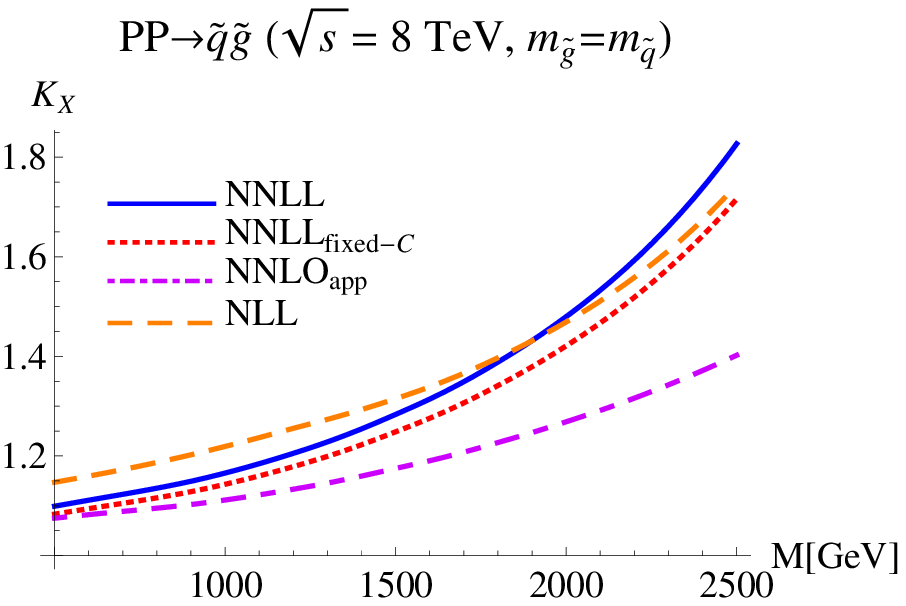}
\includegraphics[width=0.4 \linewidth]{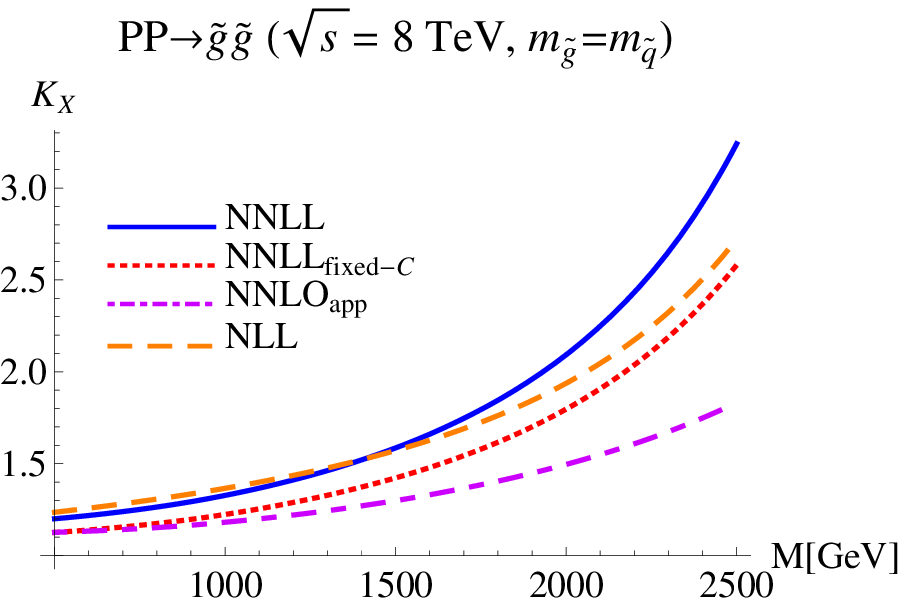}
\caption{Higher-order corrections relative to the NLO cross section
  for squark and gluino production at the LHC with $\sqrt{\hat
    s}=8$~TeV for full NNLL resummation (solid blue), NNLL with
  fixed-order Coulomb corrections (dotted red) , approximate NNLO
  (dot-dashed pink) and NLL (dashed orange). The NLL and NLO (NNLO$_{\text{app}}$ and NNLL) cross sections
  are computed with the NLO (NNLO) MSTW2008 PDFs.}
\label{fig:KNNLL}
\end{figure}

\begin{figure}[t!]
  \centering
  \includegraphics[width=0.4 \linewidth]{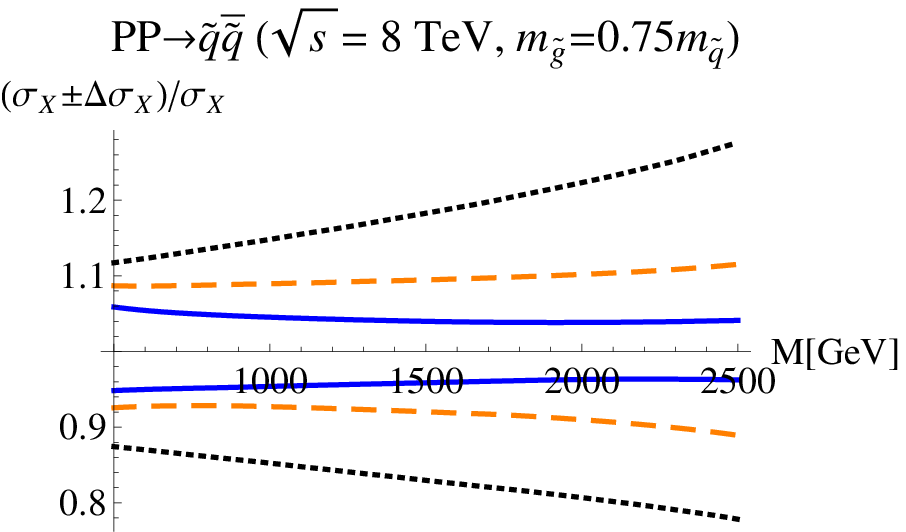}
  \includegraphics[width=0.4 \linewidth]{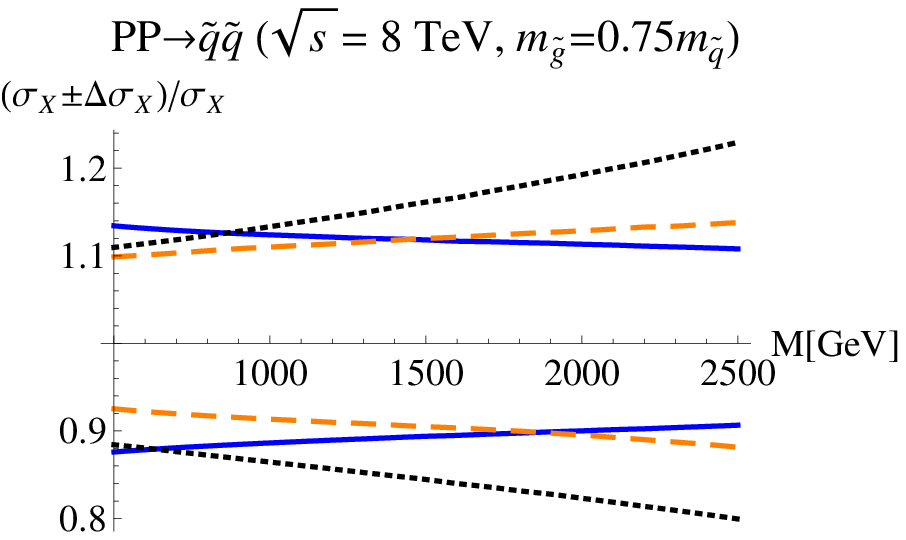}\\
 \includegraphics[width=0.4 \linewidth]{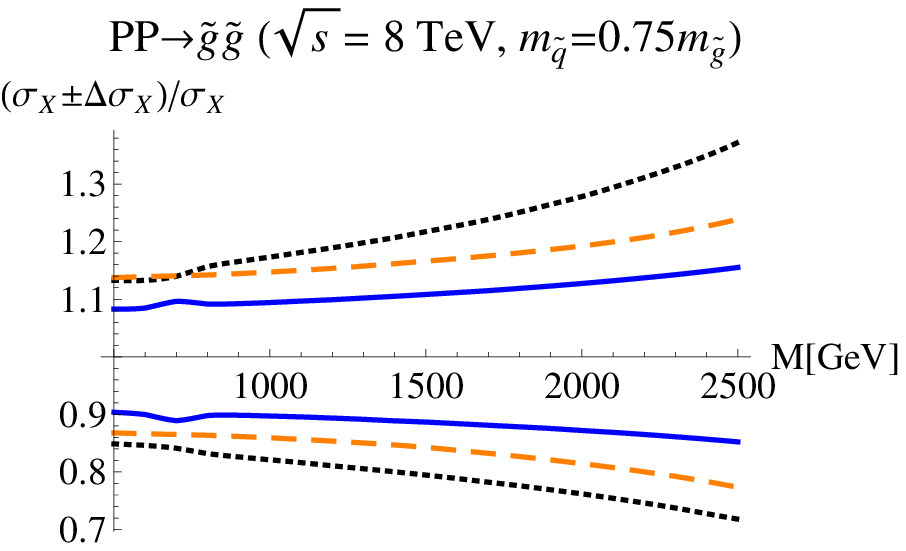}
 \includegraphics[width=0.4 \linewidth]{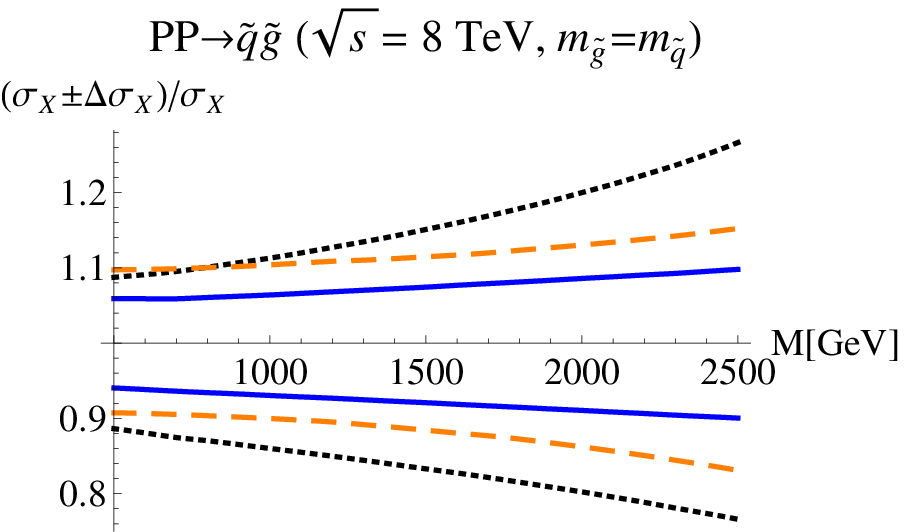}
  \caption{Total theoretical uncertainty of  the NLO approximation (dotted black), NLL (dashed orange)
and NNLL (solid blue) resummed results at the LHC with $\sqrt{s}=\,8$~TeV. All cross sections are normalized to one at the central value of the scales.}
\label{fig:uncertainty}
\end{figure}

The scale uncertainty of the NNLL predictions is estimated by varying
$\mu_f$ and $\mu_h$ as well as the scale used in the potential
function from half to twice their default values.  For the soft scale,
we employ the prescription~\cite{Beneke:2009rj} $\mu_s=k_s\, M \,\text{Max}\lbrack \beta^2
,\beta_{\text{cut}}^2\rbrack$ with $k_s=1$.  The default value of $\beta_{\text{cut}}$
is determined following~\cite{Beneke:2011mq} and the
resulting uncertainty is estimated by setting $k_s=0.5,2$ as well as
varying $\beta_{\text{cut}}$ by $\pm 20\%$ and taking the envelope of
several resummed and fixed-order approximations. As a measure of
power-suppressed terms, the non-relativistic energy $M\beta^2$ is
replaced by $E=\sqrt{\hat s}-2M$.  Finally, a constant term $\pm |
h^{R(1)}_{pp'} |^2$ is added as an estimate of unknown NNLO
corrections beyond the threshold limit.  The uncertainties from the
various sources are added in quadrature.  Our results for the
K-factors beyond NLO, $K_X=\sigma_X/\sigma_{\text{NLO}}$ with
$X=\text{NLL}, \text{NNLL}_{\text{fixed-C}}$ and NNLL, for the four
squark and gluino production processes are shown in
Figure~\ref{fig:KNNLL}.  The results show a full NNLL correction of up
to $25\%$ relative to the NLL results. The effect of Coulomb
resummation can be important in particular for
squark-antisquark and gluino-pair production.  The comparison to the
approximate NNLO results shows that corrections beyond NNLO become
sizeable beyond sparticle masses of $\sim 1.5$~TeV. The
$\text{NNLL}_{\text{fixed-C}}$ results appear to be in good agreement
with results of the Mellin-space approach to
resummation~\cite{Beenakker:2011sf} where a similar approximation is
used.  As can be seen in Figure~\ref{fig:uncertainty} the relative
uncertainty is reduced from up to $30\%$ at NLO, to at most $20\%$ at
NLL and to the $10\%$-level at NNLL.


\providecommand{\href}[2]{#2}\begingroup\raggedright\endgroup

\end{document}